\documentclass[conference,a4paper,]{IEEEtran}

\usepackage{cite}
\usepackage[T1]{fontenc}
\usepackage[latin1]{inputenc}
\usepackage{graphicx}
\usepackage{theorem}
\usepackage{amsmath}
\usepackage{amsfonts}
\ifCLASSOPTIONcompsoc
\usepackage[caption=false,font=normalsize,labelfon
t=sf,textfont=sf]{subfig}
\else
\usepackage[caption=false,font=footnotesize]{subfi
g}
\fi
\usepackage[acronym,shortcuts]{glossaries}
\usepackage{color}
\usepackage{url}
\usepackage{enumerate}
\usepackage{comment}
\usepackage{cleveref}
\usepackage{multirow}
\usepackage{color}

\usepackage{algorithm}
\usepackage[noend]{algpseudocode}
\algdef{SE}[DOWHILE]{Do}{doWhile}{\algorithmicdo}[1]{\algorithmicwhile\ #1}%

\newacronym{DFR}{DFR}{decoding failure rate}
\newacronym{ISD}{ISD}{information set decoding}
\newacronym{QC}{QC}{quasi-cyclic}
\newacronym{QC-LDPC}{QC-LDPC}{quasi-cyclic low-density parity-check}
\newacronym{QC-MDPC}{QC-MDPC}{quasi-cyclic moderate-density parity-check}
\newacronym{QC-LDGM}{QC-LDGM}{quasi-cyclic low-density generator matrix}
\newacronym{LDPC}{LDPC}{low-density parity-check}
\newacronym{LDPCC}{LDPCC}{low-density parity-check convolutional}
\newacronym{NP}{NP}{non-polynomial}
\newacronym{QD}{QD}{quasi-dyadic}
\newacronym{GRS}{GRS}{generalized Reed-Solomon}
\newacronym{AC-LDPC}{AC-LDPC}{array convolutional low-density parity-check}
\newacronym{PDC-LDPC}{PDC-LDPC}{progressive differences convolutional low-density parity-check}
\newacronym{SC-LDPC}{SC-LDPC}{spatially coupled low-density parity-check}
\newacronym{SC-LDPC-CCs}{SC-LDPC-CCs}{spatially coupled low-density parity-check convolutional codes}
\newacronym{SC-LDPC-CC}{SC-LDPC-CC}{spatially coupled low-density parity-check convolutional code}
\newacronym{AWGN}{AWGN}{additive white Gaussian noise}
\newacronym{BF}{BF}{Bit Flipping}
\newacronym{BER}{BER}{bit error rate}
\newacronym{FER}{FER}{codeword error rate}
\newacronym{TUB}{TUB}{truncated union bound}
\newacronym{BPSK}{BPSK}{binary phase shift keying}
\newacronym{SPA-LLR}{SPA-LLR}{sum-product algorithm with log-likelihood ratios}
\newacronym{RTI}{RTI}{regular time-invariant}
\newacronym{RTI-LDPCC}{RTI-LDPCC}{regular time-invariant low-density parity-check convolutional}
\newacronym{CPA}{CPA}{chosen-plaintext attack}
\newacronym{CCA2}{CCA2}{adaptive chosen-ciphertext attack}
\newacronym{SL}{SL}{security level}

{\theorembodyfont{\rmfamily}
   \newtheorem{The}{{\textbf Theorem}}[section]}
{\theorembodyfont{\rmfamily}
   }
{\theorembodyfont{\rmfamily}
   }
{\theorembodyfont{\rmfamily}
   }
{\theorembodyfont{\rmfamily}
   }
{\theorembodyfont{\rmfamily}
   }
{\theorembodyfont{\rmfamily}
   }
   
\newenvironment{proof}[1][Proof]{\textbf{#1.} }{\ \rule{0.5em}{0.5em}}

\def\0{\bar{0}}

\makeatletter
\newcommand{\tpmod}[1]{{\@displayfalse\pmod{#1}}}
\makeatother

\newcommand{\mb}[1]{#1}

\begin{document}

\title{\mb{Hindering reaction attacks by using monomial codes in the McEliece cryptosystem}}

\author{\IEEEauthorblockN{Paolo Santini$^*$, Marco Baldi, Giovanni Cancellieri and Franco Chiaraluce}
\IEEEauthorblockA{Dipartimento di Ingegneria dell'Informazione\\
Universit\`a Politecnica delle Marche\\
Ancona, Italy\\
Email: {p.santini@pm.univpm.it}, \{m.baldi, g.cancellieri, f.chiaraluce\}@univpm.it}}

\maketitle
\begin{abstract}
In this paper we study recent reaction attacks against \mb{QC-LDPC and QC-MDPC code-based cryptosystems, which allow an opponent} to recover the private parity-check matrix through its distance spectrum \mb{by observing} a sufficiently high number of decryption failures. We consider a special class of codes, known as monomial codes, to form private keys with the desirable property of having a unique and complete distance spectrum. We verify that for these codes the problem of recovering the secret key from the distance spectrum is equivalent to that of finding cliques in a graph, and use this equivalence to prove that \mb{current} reaction attacks are not applicable when codes of this type are used in the McEliece cryptosystem.
\end{abstract}


\section{Introduction}

\let\thefootnote\relax\footnotetext{$^*$The work of Paolo Santini was partially supported by Namirial S.p.A.}

Devising efficient and robust post-quantum public-key cryptosystems for key encapsulation and encryption is an important and urgent research target, as also witnessed by the recent NIST call for post-quantum cryptographic systems \cite{NISTcall2016}.
As recognized by NIST \cite{NISTreport2016}, code-based cryptosystems are among the most
promising candidates to replace systems relying on the hardness of factorizing large integers or solving discrete logarithms, like Diffie-Hellman, RSA and ElGamal, which can be broken in polynomial time using Shor's algorithm \cite{Shor1997} running on a quantum computer.
Instead, no quantum algorithm is known to quickly solve the problem of decoding a random-like 
linear block code, that hence remains a \ac{NP} time problem \cite{Berlekamp1978,May2011}.

The first code-based public-key cryptosystem was proposed by McEliece in 1978 \cite{McEliece1978},
and relied on Goppa codes \cite{Goppa1970} to form the secret key.
This yields large public keys, that is the main limitation of Goppa code-based systems.
Replacing Goppa codes with other families of more structured codes may allow reducing the public key
size; an overview of these variants can be found in \cite{Baldi2017}. 
Among them, a prominent role is played by variants based on codes with sparse parity-check matrices \cite{Baldi2012}, \cite{Misoczki2013}.
Recently, however, some new statistical attacks against such variants have been devised, exploiting Bob's reactions to gather information about the secret key \cite{Guo2016,Fabsic2017}.
These attacks exploit the fact that probabilistic iterative decoders are used, which are not bounded-distance decoders. Therefore, it may happen that decoding fails to correct some vector of intentional errors used during encryption, and in such a case Bob must communicate the decoding failure. 
In the attacks proposed in \cite{Guo2016}, \cite{Fabsic2017}, Eve impersonates Alice and sends multiple suitably chosen ciphertexts to Bob, observing his reactions.
In particular, based on Bob reactions, she estimates Bob's \ac{DFR}.
By exploiting the dependence of the \ac{DFR} on some features of the error vectors used during encryption, Eve is able to understand whether some particular binary patterns belong to the secret key or not.
By collecting this information, Eve may be able to recover a representation of the private code that allows decoding of intercepted cyphertexts.

In this paper, we consider a special family of \mb{\ac{QC-LDPC} codes}, known as monomial codes, to form the secret key.
We extend the attacks proposed in \cite{Guo2016}, \cite{Fabsic2017} to this family of codes.
We show that, as in \cite{Fabsic2017}, also in this case the secret matrix recovery problem can be translated into a graph problem.
We then introduce a special class of monomial codes to form the secret key which make \mb{all the known} reaction attacks infeasible, since they do not allow obtaining any useful information from Bob's reactions.
In particular, we show that for these codes \mb{such} attacks reduce to a brute force attack, having only one solution among a very large set of possible candidates.
\mb{This is achieved at the cost of some increase in the public key size, which is larger for monomial code-based systems with respect to classical systems based on \ac{QC-LDPC} and \ac{QC-MDPC} codes.}

\section{QC-LDPC / QC-MDPC code-based McEliece cryptosystems}\label{sec:McEliece}
Existing McEliece cryptosystem variants based on \ac{QC-LDPC}\cite{Baldi2012} and \ac{QC-MDPC}\cite{Misoczki2013} codes exploit codes with rate $R = \frac{n_0-1}{n_0}$, being $n_0$ a small integer, redundancy $p$, length $n = n_0 p$ and dimension $k = (n_0 - 1) p$.
The secret code is defined by the sparse parity-check matrix
\begin{equation}
\mathbf{H} = \left[ \mathbf{H}_{0} , \mathbf{H}_{1} , \ldots , \mathbf{H}_{n_0-1} \right],
\label{eq:HCircRow}
\end{equation}
where each block $\mathbf{H}_{i}$ is a binary circulant matrix with size $p \times p$.
The private key is formed by $\mathbf{H}$ in \eqref{eq:HCircRow} and two non-singular random matrices: a $k \times k$ scrambling matrix $\mathbf{S}$ and an $n \times n$ transformation matrix $\mathbf{Q}$.
$\mathbf{S}$ and $\mathbf{Q}$ are both formed by $p \times p$ circulant blocks, and $\mathbf{Q}$ is a sparse matrix with average row and column weight $m \ge 1$.

The \ac{QC-MDPC} code-based case is different from the \ac{QC-LDPC} code-based case in that $\mathbf{H}$ is less sparse and $\mathbf{Q}$ is reduced to an identity matrix (hence $m=1$).
Moreover, when an \ac{CCA2} secure conversion of these systems is used \cite{Bernstein2008}, $\mathbf{S}$ can also be reduced to an identity matrix.
The public key is obtained as $\mathbf{G}' = \mathbf{S}^{-1} \cdot \mathbf{G} \cdot{\mathbf{Q}^{-1}}$, where $\mathbf{G}$ is a systematic generator matrix for the code defined by $\mathbf{H}$.
Alice performs encryption by first dividing the secret message into $k$-bit blocks. 
For any of these blocks, say $\mathbf{u}$, a weight-$t$ binary error vector $\mathbf{e}$ is generated and the corresponding ciphertext block $\mathbf{x}$ is computed as
\begin{equation}
\mathbf{x} = \mathbf{u} \cdot \mathbf{G}' + \mathbf{e} = \mathbf{c} + \mathbf{e}.
\label{eq:Encryption}
\end{equation}
To perform decryption, Bob multiplies $\mathbf{x}$ by $\mathbf{Q}$ and obtains
\begin{equation}
\label{eq:Decryption}
\mathbf{x}'=\mathbf{x}\cdot\mathbf{Q}=\mathbf{u}\cdot\mathbf{S}^{-1}\cdot\mathbf{G}+\mathbf{e}\cdot\mathbf{Q}=\mathbf{u}'\cdot\mathbf{G}+\mathbf{e}',
\end{equation}
where $\mathbf{u}'\cdot\mathbf{G}=\mathbf{c}'$ is a codeword of the secret code corresponding to the information vector $\mathbf{u}'=\mathbf{u}\cdot\mathbf{S}^{-1}$, while $\mathbf{e}'=\mathbf{e}\cdot\mathbf{Q}$ plays the role of an error vector with weight $t'\leq mt$.
Bob then corrects $\mathbf{e}'$ through iterative decoding and gets $\mathbf{c}'$.
Owing to the systematic form of $\mathbf{G}$, $\mathbf{u}$ is obtained by discarding the last $p$ entries of $\mathbf{c}'$ and multiplying the remaining vector by $\mathbf{S}$.

In the \ac{QC-MDPC} code-based variant with \ac{CCA2} secure conversion, the matrices $\mathbf{S}$ and $\mathbf{Q}$ can be avoided (i.e., replaced by identity matrices); with this choice, we have $\mathbf{G}'=\mathbf{G}$.

\section{Reaction attacks\label{sec:reaction}}

The attacks introduced in \cite{Guo2016},\cite{Fabsic2017} exploit the fact that, in the \ac{QC-LDPC} and \ac{QC-MDPC} code-based systems, decryption may fail with a certain \ac{DFR}.
Moreover, the \ac{DFR} depends on the number of couples of ones that are in the same positions in $\mathbf{e}$ and in the rows of $\mathbf{H}$ and $\mathbf{Q}$ (when the latter is used).
Hence, Eve can adopt the following attack strategy:
\begin{enumerate}
\item she generates a sufficiently large number of plaintext-error vector pairs, computes the corresponding ciphertexts and sends them to Bob;
\item she classifies the produced ciphertexts into appropriate subsets, depending on the error vector structure, and waits for the corresponding decryption outcome;
\item using the observed reactions, she estimates the individual DFR over each subset. 
\end{enumerate}
Then, the DFR values of the subsets can be analyzed to obtain information about the structure of $\mathbf{H}$ (and $\mathbf{Q}$, when used).
This information can be used to reconstruct a sparse parity-check matrix $\mathbf{\hat{H}} = \mathbf{\Pi} \cdot \mathbf{H}\cdot \mathbf{Q}^T$, where $\mathbf{\Pi}$ is a $p\times p$ permutation matrix, which can be employed to efficiently decode $\mathbf{x}$ and recover $\mathbf{e}$.
Indeed, we have
\begin{align}
\mathbf{\hat{s}} & = \mathbf{\hat{H}} \cdot \mathbf{x}^T = \mathbf{\hat{H}} \cdot \mathbf{G}'^T \cdot \mathbf{u}^T + \mathbf{\hat{H}} \cdot \mathbf{e}^T = \\\nonumber
&
= \mathbf{\Pi}\cdot \mathbf{H} \cdot \mathbf{G}^T\cdot \left(\mathbf{S}^{-1}\right)^T \cdot \mathbf{u}^T+\mathbf{\hat{H}}\cdot\mathbf{e}^T =
\mathbf{\hat{H}}\cdot\mathbf{e}^T,
\end{align}
as $\mathbf{H}\cdot \mathbf{G}^T = \mathbf{0}$ by definition.
Thus, $\mathbf{\hat{s}}$ is the syndrome of $\mathbf{e}$ computed through $\mathbf{\hat{H}}$, which is a sparse matrix (since both $\mathbf{H}$ and $\mathbf{Q}$ are sparse). In the \ac{QC-MDPC} case, $\mathbf{\hat{H}}$ is just a row-permuted version of the secret parity-check matrix $\mathbf{H}$.
The opponent can then consider $\mathbf{\hat{H}}$, apply efficient syndrome decoding on $\mathbf{\hat{s}}$ and recover $\mathbf{e}$.
Since the rows of $\mathbf{\hat{H}}$ are a permuted version of those of $\mathbf{H}$, we define them as row-equivalent matrices.

The mentioned reaction attacks are based on the concept of distance spectrum, which is the set of all distances  existing between any two ones in a binary vector (cyclically closed).
Given two ones at positions $v_1$ and $v_2$, the corresponding distance is computed as
\begin{equation}
\label{eq:delta_f}
\delta(v_1,v_2)=\min{\left\{ \pm(v_1-v_2)\mod{p}\right\}}.
\end{equation}
In a circulant matrix $\mathbf{A}$, all rows produce the same distances; so, we can define its distance spectrum $\Lambda (\mathbf{A})$ as the distance spectrum of whichever row of $\mathbf{A}$ (say the first one). 
Once the distance spectrum is known, the corresponding vector can be easily recovered, apart from a cyclic permutation.
In reaction attacks, Eve first obtains the distance spectrum of the secret key, and then uses it to recover $\mathbf{\hat{H}}$.
This will be described, with reference to monomial codes, in the next section.
The recovery problem might have more than one solution, meaning that Eve might obtain more than one candidate for $\mathbf{\hat{H}}$. However, as shown in \cite{Guo2016}, \cite{Fabsic2017}, for conventional codes the number of candidates is typically small.

\section{Monomial codes in the McEliece cryptosystem}\label{sec:newvariant}
Let us consider a secret code defined by a parity-check matrix in the form
\begin{equation}
\label{eq:Hp}
\mathbf{H}=\begin{bmatrix}\mathbf{H}_{0,0}&\mathbf{H}_{0,1}&\cdots & \mathbf{H}_{0,n_0-1}\\\mathbf{H}_{1,0}&\mathbf{H}_{1,1}&\cdots & \mathbf{H}_{1,n_0-1}\\\vdots&\vdots&\ddots & \vdots \\ \mathbf{H}_{r_0-1,0}&\mathbf{H}_{r_0-1,1}&\cdots & \mathbf{H}_{r_0-1,n_0-1}
\end{bmatrix}
\end{equation}
where each block $\mathbf{H}_{i,j}$ is a $p\times p$ circulant permutation matrix, which can be represented as a monomial $x^{w_{i,j}}$ according to the classical homomorphism between $p \times p$ circulant matrices and polynomials modulo $x^p - 1$.
Thus, the secret code has length $n = n_0 p$.
It is easy to show that a matrix in the form \eqref{eq:Hp} has at least $r_0 - 1$ rows that are linearly dependent on the other rows.
Hence, the dimension of the private code is $k \ge k_0p+r_0-1$, being $k_0 = n_0 - r_0$, which usually holds with the equality sign for non-trivial choices of the matrix entries.
To avoid attacks like the one in \cite{Shooshtari2016}, we choose $p$ as a prime.
We point out that such matrices can be completely described by an $r_0\times n_0$ matrix $\mathbf{W}$, often denoted as the \textit{exponent matrix} of $\mathbf{H}$, whose elements $w_{i,j}$ are the exponents of the monomials in \eqref{eq:Hp}.

Let us suppose to use these codes, in place of those described by \eqref{eq:HCircRow}, within the schemes discussed in Section \ref{sec:McEliece}.
\mb{A generator matrix $\mathbf{G}$ can be easily obtained from \eqref{eq:Hp} in the form of $k_0 \times n_0$ circulant blocks having size $p \times p$, with the addition of $n-k-r_0p$ rows to compensate the rank deficiency of $\mathbf{H}$ in \eqref{eq:Hp} \cite{Baldi2017}.
These rows, however, do not depend on the exponents of $\mathbf{H}$. Hence, when $\mathbf{G}' = \mathbf{S}^{-1} \cdot \mathbf{G} \cdot{\mathbf{Q}^{-1}}$ is used as the public key, they can be excluded and, considering $\mathbf{G}'$ in systematic form, storing $k_0 \times r_0$ circulant blocks having size $p \times p$ requires $K_s = r_0k_0p$ bits.
}

\subsection{Reaction attacks to monomial codes}\label{sec:mono_reaction}
For matrices in the form \eqref{eq:Hp}, distances involve only ones belonging to different circulant blocks in the same row.
We point out that the values of these distances can equally be obtained considering the exponent matrix $\mathbf{W}$. 
As for the codes considered in \cite{Guo2016},\cite{Fabsic2017}, overlapping ones between $\mathbf{e}$ and one row of $\mathbf{H}$ cause a variation in the \ac{DFR}.
However, since the parity-check matrix of a monomial code contains multiple rows of circulant blocks, an opponent does not know which of them caused such an effect.

In order to understand this fact, it is useful to write the error vector in \ac{QC} form (i.e., divided into $p$-bit blocks), that is $\mathbf{e}=[\mathbf{e}_0,\mathbf{e}_1,\cdots,\mathbf{e}_{n_0-1}]$. 
Its syndrome through $\mathbf{H}$ is then $\mathbf{s}=[\mathbf{s}_0, \mathbf{s}_1,\cdots,\mathbf{s}_{r_0-1}]$, with
\begin{equation}
\label{eq:s_cmp}
\mathbf{s}_i=\sum_{j=0}^{n_0-1}{\mathbf{e}_j\mathbf{H}_{i,j}^T}.
\end{equation}
Every time $\mathbf{e}$ and (at least) one row of $\mathbf{H}$ have overlapping ones, a cancellation occurs in \eqref{eq:s_cmp}; thus, we can state that the variation in the \ac{DFR} is related to the number of such cancellations.
Since the whole error vector contributes to the computation of every block $\mathbf{s}_i$, Eve cannot know the  positions of blocks where cancellations occurred. 

In order to extend the attack in \cite{Guo2016} to monomial codes, Eve can:
\begin{enumerate}
\item define the vectors $\mathbf{a}^{(i,j)}=\left[a^{(i,j)}_0,a^{(i,j)}_1,\cdots,a^{(i,j)}_{\left\lfloor\frac{p}{2}\right\rfloor}\right]$ and $\mathbf{b}^{(i,j)}=\left[b^{(i,j)}_0,b^{(i,j)}_1,\cdots,b^{(i,j)}_{\left\lfloor\frac{p}{2}\right\rfloor}\right]$, for $i=0,\cdots,n_0-2$, $j=i+1,\cdots,n_0-1$, which are initialized with all-zero elements;
\item generate a sufficiently large set (in the order of millions, as observed in \cite{Guo2016}, \cite{Fabsic2017}) of plaintext-error vector pairs;
\item for each plaintext-error vector pair:
\begin{enumerate}
\item compute the support of $\mathbf{e}$, noted as $\Psi_{\mathbf{e}}=\left\{i_0,i_1,\cdots,i_{t-1}\right\}$;
\item encrypt the plaintext using $\mathbf{e}$ and send it to Bob;
\item for each couple $(i,j)$ of indexes in $\Psi_\mathbf{e}$, compute $z_i=\left\lfloor\frac{i}{p}\right\rfloor$,
$z_j=\left\lfloor\frac{j}{p}\right\rfloor$;
\item if $z_i\neq z_j$, compute $d=\delta(i,j)$ and increment $b^{(z_i,z_j)}_d$; in case of a decoding failure, increment $a^{(z_i,z_j)}_d$. 
\end{enumerate}
\end{enumerate}

Every time $\mathbf{e}_i$ and $\mathbf{e}_j$ have two ones at distance $d$, $b^{(z_i,z_j)}_d$ gets incremented, while $a^{(z_i,z_j)}_d$ gets incremented only in case of a decoding failure.
Then, as done in \cite{Guo2016},\cite{Fabsic2017}, the distribution of the ratios $\frac{a^{(z_i,z_j)}_d}{b^{(z_i,z_j)}_d}$ can be analyzed to guess distances in $\mathbf{W}$ (and so, in $\mathbf{H}$).
Thus, for these codes the distance spectrum $\Lambda(\mathbf{W})$ can be defined as an array formed by the sets $\lambda_{ij}(\mathbf{W})$, each one containing the $r_0$ distances between the exponents in the $i$-th and $j$-th columns.
From now on, we assume, pessimistically, that by using the described procedure Eve completely knows $\Lambda(\mathbf{W})$.

\subsection{Matrix recovery from the distance spectrum}

Let us describe how the knowledge of $\mathbf{\Lambda}(\mathbf{W})$ might be exploited to recover the structure of $\mathbf{W}$.
First of all, we point out that Eve is interested in whichever row-permuted version $\mathbf{\hat{H}}$ of $\mathbf{H}$; thus, she can simplify the problem, searching for a  $\mathbf{\hat{W}}$ with the first column made of all-zero entries.
This will be called the \textit{standard form} of the exponent matrix, and denoted as $\mathbf{W}^*$.
As done in \cite{Fabsic2017}, the matrix recovery problem can be related to the problem of finding cliques in a graph $\mathcal{G}$, which is associated to the distance spectrum and can be constructed according to Algorithm \ref{alg:constr_G}.

\begin{algorithm}[ht!]
\caption{Construction of $\mathcal{G}$}\label{alg:constr_G}
\begin{algorithmic}
\State{$\mathcal{G}\leftarrow$ graph with node $0$}
\For{$j=0,1,\cdots,n_0-1$}
\For{$d\in \lambda_{0,j}(\mathbf{W})$}
\For{$b=0,2,\cdots,2r_0-2$}
\State{$z^{(b)}_j = (j-1)p + [(p-d)\mod{p}]$}
\State{$z^{(b+1)}_j = (j-1)p + d$}
\State{Augment $\mathcal{G}$ with nodes $z^{(b)}_j$, $z^{(b+1)}_j$}
\State{Augment $\mathcal{G}$ with edges $\left( 0, z^{(b)}_j \right)$, $\left( 0, z^{(b+1)}_j \right)$}
\EndFor
\EndFor
\EndFor
     
\For{$i=1,\cdots,n_0-2$}
\For{$j=i+1,\cdots,n_0-1$}
\For{$b_i=0,1,\cdots,2r_0-1$}
\For{$b_j=0,1,\cdots,2r_0-1$}
\If{$\delta \left(z^{(b_i)}_i, z^{(b_j)}_j  \right) \in \lambda_{i,j}$}
\State{Augment $\mathcal{G}$ with edge $\left(z^{(b_i)}_i, z^{(b_j)}_j  \right)$}
\EndIf
\EndFor
\EndFor
\EndFor
\EndFor
\end{algorithmic}
\end{algorithm}
Each $n_0$-clique in $\mathcal{G}$ containing the node 0 represents a possible solution for one row of $\mathbf{W}^*$, in the sense that the corresponding distances are compliant with $\Lambda(\mathbf{W})$. 
We point out that, for every clique $\Gamma=\left\{\gamma_0,\gamma_1,\cdots,\gamma_{n_0-1}\right\}$, the graph contains also the clique $\Gamma^*$, with elements $\gamma_i^*  = p \left\lfloor \frac{\gamma_i}{p} \right\rfloor + \left[\left( p-\gamma_i\right) \mod{p} \right]$.
This property can be easily proven by verifying that the application of \eqref{eq:delta_f} to every couple of nodes in $\Gamma$ and $\Gamma^*$ produces the same set of distances; thus, since every row in $\mathbf{W}^*$ corresponds to two cliques in $\mathcal{G}$, the number of $n_0$-cliques in the graph  cannot be lower than $2r_0$.

\subsection{Monomial codes with identical distance spectra\label{sec:unif_mono_proc}}

Let us introduce a special class of monomial codes, designed with the goal of maximizing the number of candidates which can be obtained by the matrix recovery procedure.
Basically, the idea is to carefully choose the exponents of the monomials in $\mathbf{H}$, in order to obtain a distance spectrum $\Lambda(\mathbf{W})$ such that
\begin{equation}
\label{eq:DS_ind}
\lambda_{i,j}(\mathbf{W})=\left\{ 0,1,2,\cdots,\left\lfloor \frac{p}{2}\right\rfloor \right\}\hspace{2mm}\forall i,j.
\end{equation}
A matrix with this feature can be obtained by the procedure reported next. The resulting matrix has $r_0 = \left\lceil\frac{p}{2}\right\rceil$, $n_0 = p$ and \mb{a constant row weight equal to $n_0 = \sqrt{n}$, which is typical of \ac{QC-MDPC} codes.}

\subsection*{Exponent matrix construction}
\begin{enumerate}
\item Randomly pick $\mathbf{y}=\left[ y_0, y_1, \cdots , y_{r_0-1} \right]$, where $y_i$ is an integer $\in [0; p-1]$; 
\item randomly pick a permutation $\mathbf{v}=\left[v_0,v_1,\cdots,v_{p-1} \right]$ of the vector $\left[0,1,\cdots, p-1 \right]$;
\item randomly pick a permutation $\mathbf{q}=\left[ q_0,q_1,\cdots,q_{r_0-1} \right]$ of the vector $\left[ 0,1,\cdots, \left\lfloor\frac{p}{2}\right\rfloor \right]$;
\item for $i=0,1,\cdots,n_0-1$, compute the $i$-th column of $\mathbf{W}$ as
\begin{equation}
\label{eq:col_WH}
\mathbf{y}^T+v_i\mathbf{q}^T\mod{p}.
\end{equation}
\end{enumerate}

\begin{The}{Let $\mathbf{W}$ be a matrix constructed according to the above procedure; then,  \eqref{eq:DS_ind} holds.}
\end{The}

\begin{proof}
The image of $\delta(i,j)$ contains the $\left\lceil \frac{p}{2}\right\rceil$ integers in the range $\left[ 0 ; \left\lfloor \frac{p}{2}\right\rfloor \right]$; since each set $\lambda_{i,j}(\mathbf{W})$ contains $r_0 = \left\lceil \frac{p}{2}\right\rceil$ elements, to prove the theorem it is sufficient to demonstrate that, for each couple of columns $(i,j)$, no duplicated distances can exist in $\lambda_{i,j}(\mathbf{W})$.
Let us consider two columns, identified by the indexes $i$ and $j$, and two rows, identified by the indexes $l$ and $m$; the corresponding exponents in $\mathbf{W}$ will be $w_{l,i}$, $w_{l,j}$, $w_{m,i}$ and $w_{m,j}$.
Because of the structure of $\mathbf{W}$, we have
\begin{center}
$w_{l,i} = y_l + v_i q_l\mod{p}$,
\\$w_{l,j} = y_l + v_j q_l\mod{p}$,
\\$w_{m,i} = y_m + v_i q_m\mod{p}$,
\\$w_{m,j} = y_m + v_j q_m\mod{p}$.
\end{center}
Let us consider the exponents in the $l$-th row: we have   
$w_{l,i}-w_{l,j} \equiv (v_i-v_j) q_l \equiv \Delta_{i,j} q_l \mod{p}$. 
For the $m$-th row, in the same way, we have $w_{m,i}-w_{m,j} \equiv \Delta_{i,j} q_m\mod{p}$.
Since $\mathbf{v}$ is a permutation of the integers from $0$ to $p-1$, it is easy to see that $\Delta_{i,j} \in [-p+1; -1] \cup [1; p-1]$, so $\Delta_{i,j} \not\equiv 0 \mod{p}$.
We must now prove that $\delta(w_{l,i},w_{l,j})\neq \delta(w_{m,i},w_{m,j})$, i.e.,
\begin{equation}
\label{eq:eqThe}
\min{\left\{\pm \Delta_{i,j} q_l \mod{p}\right\}}\neq\min{\left\{\pm \Delta_{i,j} q_m \mod{p}\right\}}.
\end{equation}
First of all, $\pm \Delta_{i,j} q_l \equiv \pm \Delta_{i,j} q_m \mod{p}$ can only be satisfied if $q_l \equiv q_m \mod{p}$, which is not possible since the values in $\mathbf{q}$ are all distinct and smaller than $p$.
So, we must consider the case of $\pm \Delta_{i,j} q_l \equiv \mp \Delta_{i,j} q_m \mod{p}$, which gives $\pm \Delta_{i,j} (q_l+q_m) \equiv 0 \mod{p}$.
This relation cannot be satisfied as well, since $0 < q_l + q_m < 2\left\lfloor \frac{p}{2} \right\rfloor= p-1$.
\end{proof}

It can be easily proven that an exponent matrix satisfying \eqref{eq:DS_ind} is associated to a graph having $p^{n_0-1}$ cliques of size $n_0$; it can also be proven that this number corresponds to the maximum number of cliques which can exist in a graph constructed according to Algorithm \ref{alg:constr_G}.
In addition, we must consider that all the matrices constructed according to the above procedure have the same distance spectrum, and thus share the same graph $\mathcal{G}$.
Hence, performing a reaction attack like the one described in section \ref{sec:mono_reaction} is pointless: the distance spectrum is no longer secret, but there is no way to distinguish among all the possible candidates for $\mathbf{W}^*$.
Thus, the reaction attack is reduced to a brute-force attack: Eve keeps on generating and testing matrices in the form of $\mathbf{W}^*$, until a valid one is found out.
This search is facilitated by the fact that each row-permuted version of $\mathbf{W}^*$ is acceptable: indeed, the $i$-th column of $\mathbf{W}^{*}$ can be written as:
\begin{align}
\label{eq:std_form}
\mathbf{w}^{*T}_i&=\mathbf{w}^{T}_0+v_i\mathbf{q}^T-\mathbf{w}^{T}_0-v_0\mathbf{q}^T\mod{p}= \\\nonumber
& = \left( v_i-v_0 \right)\mathbf{q}^T\mod{p} = \\\nonumber
& = v_i^* \mathbf{q}^{T}\mod{p},
\end{align}
where $\mathbf{w}_0^T$ is the first column of $\mathbf{W}$ and $v_i^* = v_i-v_0\mod{p}$ (so $v_0^* = 0$). 
The vector $\mathbf{v}^*=\left[ v^*_0, v^*_1, \cdots, v^*_{n_0-1}\right]$ corresponds to a permutation of the integers $\left\{ 0,1,\cdots, p-1 \right\}$, having $0$ as its first element.
Eve is just looking for a matrix $\mathbf{\hat{W}}^*$ which is row-equivalent to $\mathbf{W}^*$, so she can fix the order of the elements of $\mathbf{q}$ and only try different configurations of $\mathbf{v}^*$.
Since there are $(p-1)!$ possible configurations for $\mathbf{v}^*$, this means that there are $N_W=(p-1)!$ possibilities for $\mathbf{\hat{W}}^*$.
We point out that this number is very large even for small values of $p$: as an example, for $p = 26$ we have $(p-1)! \approx 2^{83.7}$.

To conclude our analysis, we must consider the occurrence of two different row-equivalent matrices $\mathbf{W}^{*}$: in this case, also the corresponding parity-check matrices would be row-equivalent and could be used to decode the same code.
However, by the following theorems we prove that such a case cannot occur; thus, once a pair of private-public keys is generated, there can only be one matrix $\mathbf{W}^*$ allowing decoding of intercepted ciphertexts.

\begin{The}\label{the:perm_sets}{ Let $p$ be a prime, and let $\mathbf{z}=\left[z_0,z_1,\cdots,z_{p-2} \right]$ be a permutation of the integers in the range $[1; p-1]$; then, the sets $\Im(\alpha, \mathbf{z}) = \left\{ \alpha z_i \hspace{1mm} | \hspace{1mm} i=0,1,\cdots,p-2\right\}$, for $\alpha = 1,2,\cdots,p-1$,
are all distinct permutations of the integers in the range $[1; p-1]$.}
\end{The}

\begin{proof}
Based on combinatorial arguments and omitted due to lack of space. 
\end{proof}

\begin{The}\label{the:primes_prop}{ Let $p$ be a prime, and let $\alpha$ be an integer such that $2\leq \alpha \leq \left\lfloor \frac{p}{2}\right\rfloor$; then, there always exists an integer $\beta$ such that $2\leq \beta \leq \left\lfloor \frac{p}{2}\right\rfloor$ and $\alpha\beta \mod{p} > \left\lfloor \frac{p}{2}\right\rfloor $.}
\end{The}

\begin{proof}
Based on combinatorial arguments and omitted due to lack of space.
\end{proof}

\begin{The}\label{the:row_equiv}{ Let $\mathbf{W}^{(0)}$ and $\mathbf{W}^{(1)}$ be two exponent matrices generated according to the previous procedure, with $\mathbf{v}^{(0)}\neq \mathbf{v}^{(1)}$,
and let $\mathbf{W}^{*(0)}$ and $\mathbf{W}^{*(1)}$ denote their corresponding standard forms.
Then, $\mathbf{W}^{*(0)}$ and $\mathbf{W}^{*(1)}$ cannot be row-equivalent.}
\end{The}
\begin{proof}
Extending \eqref{eq:std_form} to the rows, those at position $i$ in the matrices $\mathbf{W}^{*(b)}$, for $b=0,1$, can be expressed as $\mathbf{w}_i^{*(b)} = q_i^{(b)}\mathbf{v}^{*(b)}\mod{p}$.
We can impose $q_0^{(0)}=q_0^{(1)}=0$ and, since $v^{*(0)}_0 = v^{*(0)}_1 = 0$, we have
\begin{equation}
\mathbf{W}^{*(b)} = \begin{bmatrix}0 & \begin{matrix}\cdots & 0\end{matrix}\\\begin{matrix}\vdots\\0\end{matrix} & \mathbf{\tilde{W}}^{*(b)}
\end{bmatrix}.
\end{equation}
where $\mathbf{\tilde{W}}^{(b)}$ is an $(r_0-1) \times (n_0-1)$ matrix, whose $i$-th row is defined as $q^{(b)}_i\left[ v^{*(b)}_1,v^{*(b)}_2,\cdots,v^{*(b)}_{p-1} \right] \mod{p} = q^{(b)}_i \mathbf{\tilde{v}}^{(b)} \mod{p} = \Im\left(q_i^{(b)}, \mathbf{\tilde{v}}^{(b)} \right)$. 
If $\mathbf{\tilde{W}}^{*(0)}$ and $\mathbf{\tilde{W}}^{*(1)}$ are row-equivalent, then each row of the former matrix must correspond to a distinct row of the latter matrix. Because of Theorem \ref{the:perm_sets}, this means that there must be a bijection $\Phi$, having domain and codomain equal to $D=\left\{1,2,\cdots,\left\lfloor \frac{p}{2}\right\rfloor \right\}$,  such that
\begin{equation}
\label{eq:ide_1}
q_i^{(0)}\mathbf{\tilde{v}}^{(0)} = \Phi\left(q_i^{(0)} \right) \mathbf{\tilde{v}}^{(1)}\mod{p}, \hspace{1mm} \forall i \in [1 ; r_0-1].
\end{equation}
In particular, let us suppose that for $i=\alpha$ we have $q_{\alpha}^{(0)}=1$; then, $\mathbf{\tilde{v}}^{(0)} = \Phi(1) \mathbf{\tilde{v}}^{(1)}\mod{p}$.
Obviously, the case of $\Phi(1)=1$ is not allowed, since it means that $\mathbf{\tilde{v}}^{(0)}=\mathbf{\tilde{v}}^{(1)}$.  
By substitution into \eqref{eq:ide_1}, we obtain
\begin{equation}
q_i^{(0)} \Phi(1) \mod{p}= \Phi\left(q_i^{(0)} \right), \forall i \in [1 ; r_0-1].
\end{equation}
Because of Theorem \ref{the:primes_prop}, there will always be (at least) one value of $i$, say $i'$, such that $\Phi\left(q_{i'}^{(0)} \right) > \left\lfloor \frac{p}{2} \right\rfloor$, thus proving that such a bijection $\Phi$ cannot exist.
\end{proof}

\subsection{System parameters design}
Being resistant to \mb{all the known} reaction attacks, the parameters of the proposed cryptosystem must be designed taking into account all the other known attacks.
Among them, the most dangerous ones are those based on \ac{ISD}, whose complexity can be estimated through the analysis in \cite{CantoTorres2016_bis}: given a code with length $n$, dimension $k$ and minimum distance $w$, the complexity of \ac{ISD} can be computed as $C_{\texttt{ISD}}(n,k,w)\approx 2^{-cw}$, with $c=\log_2{\left( 1-\frac{k}{n}\right)}$.

In the case of a key recovery attack, an \ac{ISD} algorithm can be used to search for low-weight codewords in the dual of the public code, which admits $\mathbf{H}$ as a valid generator matrix. Indeed, because of its sparsity, the probability that the sum of two (or more) of its rows (having weight $n_0$) results in a vector with weight $\leq n_0$ is negligible, so the rows of $\mathbf{H}$ can be considered as the minimum weight codewords of the dual code.
Since there are $r_0$ circulant block rows in $\mathbf{H}$, the opponent needs to determine $r_0$ minimum weight codewords of the dual code.
An \ac{ISD} algorithm can also be used to correct the intentional error vector and perform decoding of the public code.
Thus, taking into account the speed-up due to the \ac{QC} nature of the code as in \cite{Misoczki2013}, the work factor (WF) of \ac{ISD}-based attacks can be estimated as $WF_{\texttt{KR}}=\frac{r_0}{p}C_{\texttt{ISD}}(n,n-k,n_0)$ for key recovery and $WF_{\texttt{DA}}=\frac{1}{\sqrt{p}}C_{\texttt{ISD}}(n,k,t)$ for decoding.

Considering these expressions, we have designed three system instances, for as many different \ac{SL} values. 
The theoretical bit flipping threshold estimation \cite{Baldi2012} has been used to predict the error correcting capability of the codes.
The parameters of these instances are shown in Table~\ref{tab:sys_parameters}.

\begin{table}[!t]
\caption{System parameters for different security levels
\label{tab:sys_parameters}}
\centering
\begin{tabular}{|c|c|c|c|c|c|c|}
\hline
$\mathbf{SL}$ & $\mathbf{p}$ & $\mathbf{n_0}$ & $\mathbf{r_0}$ & $\mathbf{t}$ & $\mathbf{N_W}$ & $K_s$(kB)\\\hline\hline
80 & 103 & 103 & 52 & 84 & $2^{538}$ & 34.14     \\\hline
128 & 137  & 137 & 69 & 132 & $2^{773}$ & 80.36 \\\hline
256 & 257 & 257 & 129 & 261 & $2^{1684}$ & 530.45 \\\hline
\end{tabular}
\end{table}

\section{Conclusions}
We have proposed a \mb{public-key cryptosystem based on monomial codes, and considered a special class of such codes that are robust against state-of-the-art reaction attacks exploiting decoding failures.} 
To the best of our knowledge, this is the first McEliece cryptoystem variant that \mb{admits non negligible \ac{DFR} (i.e., $>2^{-\mathrm{SL}}$) and, at the same time, hinders attacks of this type.}
Even though the public keys are larger than in classical \ac{QC-LDPC} and \ac{QC-MDPC} code-based variants (which, however, are vulnerable to reaction attacks), they remain significantly smaller than those of Goppa code-based variants, which use bounded-distance decoders and are not subject to reaction attacks.
For instance, a Goppa code-based system achieving $\mathrm{SL} = 80$ bits would require a systematic public key of \mb{about $57$ kB \cite{Bernstein2008}}, which is much larger than \mb{that of the proposed system instance with the same \ac{SL}}.

\bibliographystyle{IEEEtran}
\bibliography{Archive}

\end{document}